\input epsfig.sty
\input suprtab.sty
\documentstyle[editedvolume]{crckapb} 

\begin{opening}
\title{High Energy Gamma-Ray Emission from Blazars:}
\subtitle{EGRET OBSERVATIONS\footnote{Invited review paper to appear in {\it Observational 
Evidence for Black Holes in the Universe}, ed. S.~K. Chakrabarti, 1999 (Dordrecht: Kluwer) pg 215-230.}}

\author{R. MUKHERJEE}
\institute{Barnard College \& Columbia University\\
           Dept. of Physics \& Astronomy\\
           3009 Broadway, 506 Altschul\\
           New York, NY 10027}

\end{opening}

\begin{document}

\begin{abstract}
We will present a summary of the observations of blazars by the 
Energetic Gamma Ray Experiment Telescope (EGRET) 
on the Compton Gamma Ray Observatory (CGRO).  EGRET has detected high 
energy $\gamma$-ray emission at energies greater than 100 MeV 
from more that 50 blazars. These sources show inferred isotropic luminosities 
as large as $3\times 10^{49}$ ergs s$^{-1}$. One of the most remarkable 
characteristics of the EGRET observations is that the $\gamma$-ray luminosity 
often dominates the bolometric power of the blazar. A few of the blazars 
are seen to exhibit variability on very short time-scales of one day or 
less. The combination of high luminosities and time variations seen in the 
$\gamma$-ray data indicate that $\gamma$-rays are an important component of the 
relativistic jet thought to characterize blazars. Currently most models for 
blazars involve a beaming scenario. In leptonic models, where electrons are 
the primary accelerated particles, $\gamma$-ray emission is believed to be 
due to inverse Compton scattering of low energy photons, although opinions 
differ as to the source of the soft photons. Hardronic models involve 
secondary production or photomeson production followed by pair cascades, and 
predict associated neutrino production. 
\end{abstract}

\section{Introduction}

One of the most striking accomplishments of the Energetic Gamma Ray 
Experiment Telescope (EGRET) instrument on 
the Compton Gamma-Ray Observatory (CGRO) is the detection 
of high-energy $\gamma$-rays from active galaxies whose emission at 
most wavebands is dominated by non-thermal processes. These objects, 
called ``blazars,'' are highly variable at most frequencies and are 
bright radio sources. Prior to the launch of CGRO, 3C 273, discovered by 
COS-B (Swanenburg et al. 1978), was the only known extragalactic source 
of $\gamma$-rays. Since then, EGRET has detected more than 50 blazars in 
high energy ($>100$ MeV) $\gamma$-rays (Mukherjee et al. 1997; Thompson 
et al. 1995; 1996). 

The blazars detected by EGRET all share the common characteristic that they 
are radio-loud, flat-spectrum radio sources, with radio spectral indices 
$\alpha_r\geq -0.6$ (von Montigny et al. 1995). 
Several of these blazars are known to demonstrate superluminal motion 
of components resolved with VLBI (3C 279, 3C 273, 3C 454.3, PKS 0528+134, 
for example). The blazar class of active galactic nuclei 
(AGN) includes BL Lac objects, highly polarized quasars (HPQ), or optically 
violent variable (OVV) quasars and are 
characterized by one or more of the properties of this source class, namely, 
a non-thermal continuum spectrum, a flat radio spectrum, strong variability 
and optical polarization. 
For many of the EGRET-detected blazars, the $\gamma$-ray energy flux is dominant 
over the flux in lower energy bands. The redshifts of these sources range from 
0.03 to 2.28 and the average photon spectral index, assuming a simple power 
law fit to the spectrum, is $\sim 2.2$.
Many of the blazars exhibit variability in their $\gamma$-ray flux on 
timescales of several days to months. In addition, blazars exhibit strong 
and rapid variability in both optical and radio wavelengths. 

Of the 51 blazars reviewed here, 14 are BL Lac objects, and the rest are 
flat spectrum radio quasars (FSRQs). BL Lac objects generally have stronger 
polarization and weaker optical lines. In fact, some BL Lac objects have no 
redshift determination because they have no identified lines above their 
optical continuum. FSRQs are generally more distant and more luminous 
compared to the BL Lac objects. 

This review summarizes the present knowledge on $\gamma$-ray observations of 
blazars by EGRET. A brief description of the EGRET instrument and data 
analysis techniques, and the list of blazars detected by EGRET is given 
in \S 2. Temporal variations and $\gamma$-ray luminosity of blazars are 
discussed in \S\S 3 \& 4. Section 5 describes the spectral energy distribution 
of blazars and summarizes the various models that have been proposed to 
explain the $\gamma$-ray emission in blazars.  

\section {EGRET observations and analysis}

\subsection{The EGRET Instrument}
EGRET is a $\gamma$-ray telescope that is sensitive in the 
energy range $\sim$ 30 MeV to 30 GeV. It has the standard components of 
a high-energy $\gamma$-ray instrument: an anticoincidence dome to discriminate 
against charged particles, a spark chamber particle track detector with 
interspersed high-$Z$ material to convert the $\gamma$-rays into 
electron-positron pairs, a triggering telescope to detect the presence 
of the pair with the correct direction of motion, and an energy measurement 
system, which in the case of EGRET is a NaI(Tl) crystal. EGRET has an 
effective area of 1500 cm$^2$ in the energy range 0.2 GeV to 1 GeV, decreasing 
to about one-half the on-axis value at $18^\circ$ off-axis and to one-sixth 
at $30^\circ$. The instrument is described in details 
by Hughes et al. (1980) and Kanbach et al. (1988, 1989) and the preflight and 
postflight calibrations are given by Thompson et al. (1993) and 
Esposito et al. (1998), respectively. 

Although EGRET records individual photons in the energy range 30 MeV to 
about 30 GeV, there are several instrumental characteristics that limit 
the energy range for which time variation investigations of blazars are 
viable. At the low end of the energy range, below $\sim 70$ MeV, 
there are systematic uncertainties that make the spectral information 
marginally useful. In addition, the deteriorating point spread function 
(PSF) and energy resolution at low energies, make analysis more difficult. 
At high energies, although the systematic uncertainties are reduced, and the 
PSF and energy resolution are more reasonable, because of the steeply 
falling spectra, few photons are detected above 5 GeV. 

The angular resolution of EGRET is energy dependent, varying from about 
$8^\circ$ at 60 MeV to $0.4^\circ$ above 3 GeV (68\% containment). The 
positions of sources are detected with varying accuracy: better than 
$0.1^\circ$ for the very bright sources, or at least 0.5$^\circ$ for sources 
just above the detection threshold. 

The threshold sensitivity of EGRET ($> 100$ MeV) for a single observation is 
$\sim 3\times 10^{-7}$ photons cm$^{-2}$ s$^{-1}$, and is only about a factor 
of 50-100 below the maximum blazar flux ever observed. The dynamic range 
for most observations of blazar variations is, therefore, fairly small. 

\subsection{EGRET Data Analysis}

The blazars described here were 
typically observed by EGRET for a period of 1 to 2 weeks; however, several of 
them were observed for 3 to 5.5 weeks. Following the standard EGRET 
processing of individual $\gamma$-ray events, summary event files were produced 
with $\gamma$-ray arrival times, directions and energies. For the observations 
reported here, photons coming from directions greater than $30^\circ$ from 
the center of the field of view (FOV) were not used, in order to restrict the 
analysis to photons with the best energy and position determinations. In 
addition, exposure history files were produced containing information on the 
instrument's mode of operation and pointing. These maps were used to generate 
skymaps of counts and intensity for the entire field of view for each 
observation, using a grid of $0.5^\circ\times 0.5^\circ$. The intensity maps 
were derived simply by dividing the counts by the exposure. The EGRET data 
processing techniques are described further by Bertsch et al. (1989). 

The number of source photons, distributed according 
to the instrument PSF in excess of the diffuse 
background, was optimized. An $E^{-2}$ photon spectrum was initially assumed 
for the source search. The background diffuse radiation was taken to be a 
combination of a Galactic component caused by cosmic ray interactions in 
atomic and molecular hydrogen gas (Hunter et al. 1997), as well as an 
almost uniformly distributed component that is believed to be of 
extragalactic origin (Sreekumar et al. 1998). 

The data were analyzed using the method of maximum 
likelihood as described by Mattox et al. (1996) and Esposito et al. (1998). 
The likelihood value, $L,$ for a model of the number of $\gamma$-rays in each 
pixel of a region of the map is given by the product of the probability that 
the measured counts are consistent with the model counts assuming a Poisson 
distribution. The probability of one model with likelihood, $L_1,$ better 
representing the data than another model with likelihood, 
$L_2,$ is determined from twice the difference of the logarithms of the 
likelihoods, $2(\ln L_2-\ln L_1).$ This difference, referred to as the test 
statistic $TS$, is distributed like $\chi^2$ with the number of degrees of 
freedom being the difference in the number of free parameters in the two 
models. The flux of the point source and the flux of the diffuse background 
emission in the model are adjusted to maximize the likelihood. The 
significance of a source 
detection in sigma is given approximately by the square root of $TS$. 

\subsection{EGRET Observations}

The 51 blazars listed in Table 1 were all detected by EGRET above 100 MeV 
during the period of EGRET observations from 1991 April to 1995 
September (Phases 1 through 4 of CGRO) (Mukherjee et al. 1997). 
Some of these blazar associations are not certain; 
Mattox et al. (1997a) find only 42 identifications to have high confidence. 
Conversely, some of the unidentified high-latitude EGRET sources are likely 
to be blazars. In addition to the 42 considered strongest, Mattox et al. 
(1997a) note 16 possible associations with bright flat-spectrum, blazar-like 
radio sources. Typically, each blazar listed in Table 1 was seen in several 
different viewing periods (VPs). The maximum and minimum fluxes observed 
for each blazar is indicated in Table 1. A more complete list of blazar 
detections by EGRET may be found in the third EGRET catalog 
(Hartman et al. 1998). 

\begin{figure}%
\centerline{\epsfig{file=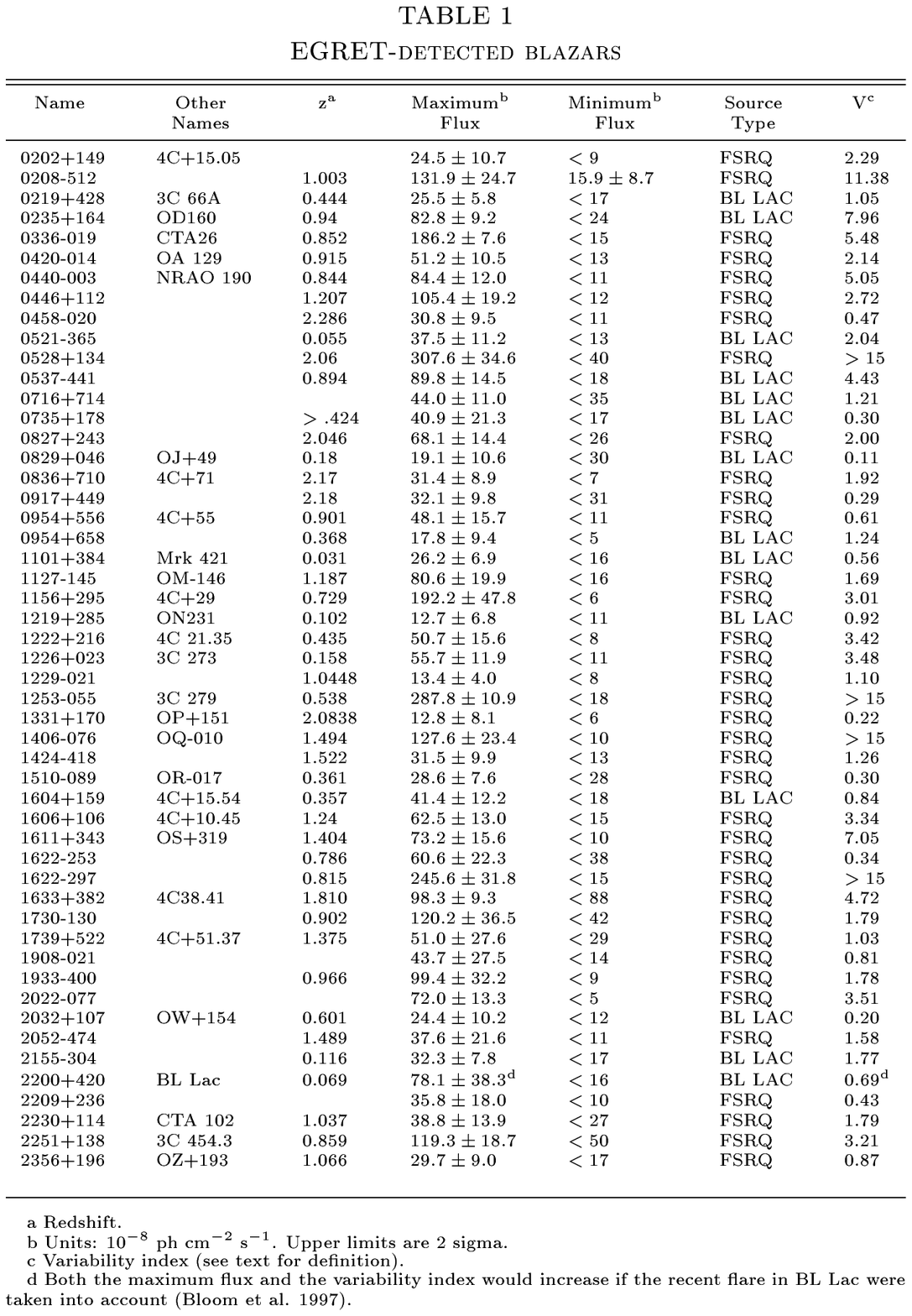,height=7.4in,width=5.3in,bbllx=50pt,bblly=130pt,bburx=500pt,bbury=680pt,clip=.}}
\end{figure}
\setcounter{figure}{0}
 
\section{Time variability}
The fluxes of the blazars detected by EGRET have been found to be variable on 
time scales of a year or more down to well under a day. Long term variations 
of blazars have been addressed earlier by several authors (eg. von Montigny 
et al. 1995; Hartman et al. 1996a; Mukherjee et al. 1997). 
In some cases, many of the detected blazars have 
exhibited flux variations up to a factor of about 30 between different 
observations. Figure 1 shows the flux history of four 
EGRET-detected blazars. The horizontal bars on the individual data 
points denote the extent of the VP for that observation. 
Fluxes have been plotted for all detections greater than $2\sigma$. 
For detections below $2\sigma$, upper limits at the 95\% confidence level 
are shown. A systematic uncertainty of 6\% was added in quadrature with the 
statistical uncertainty for each flux value, consistent with the analysis 
of McLaughlin et al. (1996) on EGRET source variability. 

\begin{figure}[t!] 
\centerline{\epsfig{file=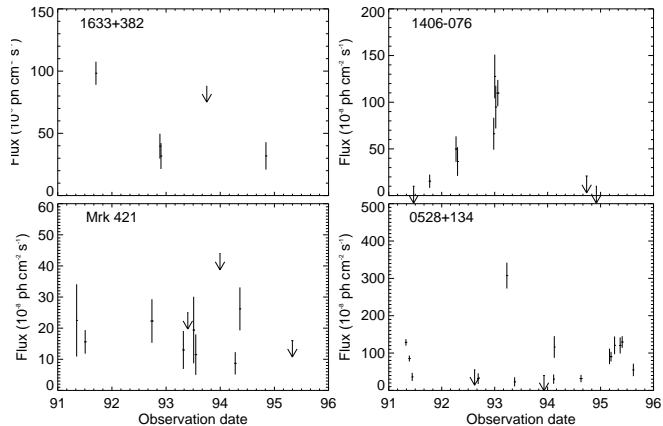,height=2.5in,bbllx=20pt,bblly=350pt,bburx=600pt,bbury=750pt,clip=.}}
\vspace{10pt}
\caption{Flux history of four blazars (PKS 1633+382, PKS 1406-076, Mrk 421, and PKS 0528+134) from 1991 April to 1995 September, as detected by EGRET. $2\sigma$ upper limits are indicated by downward arrows.}
\label{fig1}
\end{figure}
 
In order to quantify the flux variability of the blazars in Table 1, 
Mukherjee et al. (1997) calculated the variability index, $V=\log Q$, as 
defined by McLaughlin et al. (1996), where $Q=1-P_{\chi}(\chi^2,\nu).$ 
Here $P_\chi(\chi^2,\nu)$ is the probability of observing $\chi^2$ or 
something larger from a $\chi^2$ distribution with $\nu$ degrees of freedom. 
The flux versus time data were fit to a constant flux and the reduced 
$\chi^2_\nu$, for $\nu$ degrees of freedom, was calculated using the least 
square fit method. For a nonvariable source, a 
constant flux is expected to fit the data well, and the mean value of the 
$\chi^2$ distribution is expected to be equal to the number of degrees of 
freedom in the data. The quantity $V$ is used to 
judge the strength of the evidence for flux variability. 
Following the classification of 
McLaughlin et al. , $V<0.5$ was taken to indicate non-variability, $V\geq 1$ to indicate 
variability, and $0.5 \leq V < 1$ as uncertain. Table 1 lists the value of 
$V$ for each source. 

Of the 51 blazars reviewed here, 35 are found to be variable ($V\geq 1$), 
9 are non-variable ($V< 0.5$), and 7 fall in the range of uncertain 
variability. It should be noted that, although the criterion used here to 
gauge variability of a blazar is somewhat arbitrary, it does provide a way to 
compare the numbers obtained. Also, as McLaughlin et al.  (1996) have noted, 
changing these criteria by 20\% yields similar results. If the FSRQs 
and the BL Lac objects are considered separately, it is found that 
76\% of the FSRQs in the sample are variable, while 16\% are non-variable. 
Similarly, for the BL Lac objects, 50\% are variable, while 21\% are 
definitely non-variable. It should be noted that the low intrinsic luminosity 
of BL Lac objects could bias observations (see discussions in \S 5.2). 
Figure 2 shows the distribution of 
the variability indices for the FSRQs and the BL Lac objects. The BL Lac 
objects in the data set are found to be less variable on the average than 
the FSRQs. Recent observations of flares in BL Lac objects, however, modify 
some of these conclusions. For example, BL Lac was detected during a 
$\gamma$-ray outburst with an average flux of $(171\pm42)\times 10^{-8}$ 
photons cm$^{-2}$ s$^{-1}$ in July 1997 (Bloom et al. 1997). BL Lac would have 
a high value of $V$ in Table 1, if this information were taken into account. 

\begin{figure}[t!] 
\centerline{\epsfig{file=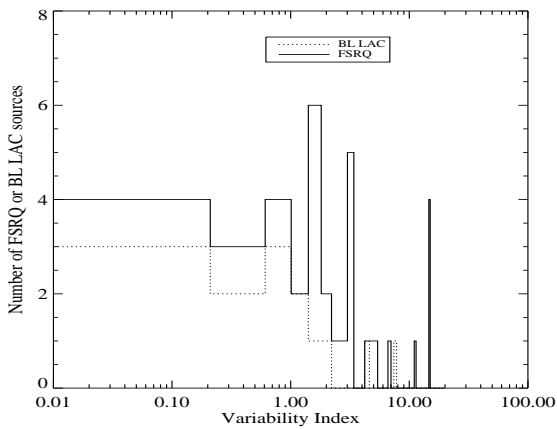,height=2.5in,width=3in,bbllx=70pt,bblly=100pt,bburx=580pt,bbury=670pt,clip=.}}
\vspace{10pt}
\caption{Distributions of the variability indices for BL Lac objects and FSRQs.}
\label{fig2}
\end{figure}
 
Figure 3 shows a plot of the variability index as a 
function of the weighted average flux for the blazars in Table 1. Note that 
the sources that have the highest average fluxes all have high variability 
indices. Only 2 out of 18 blazars with average flux less than 
$1\times 10^{-7}$ photons cm$^{-2}$ s$^{-1}$ have a variability index greater 
than 2.5. In fact, there are no non-variable blazars in the sample that have 
high flux. 

\begin{figure}[t!] 
\centerline{\epsfig{file=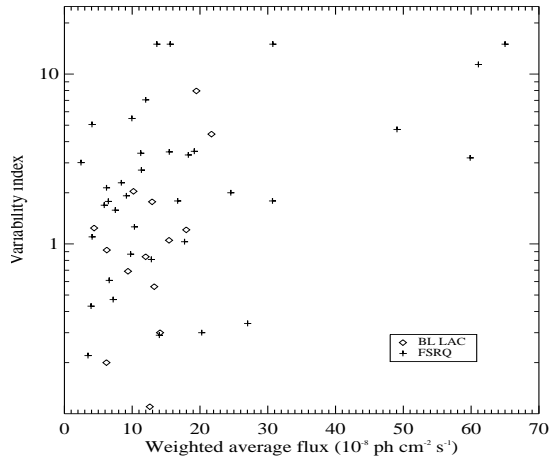,height=2.7in,width=3in,bbllx=70pt,bblly=100pt,bburx=580pt,bbury=670pt,clip=.}}
\vspace{10pt}
\caption{Variability index plotted as a function of the weighted average flux for BL Lac objects and FSRQs. The BL Lac objects 
are denoted by diamonds and the FSRQs by plus signs in the figure. }
\label{fig3}
\end{figure}
 
The study of short term variability in blazars is always limited by the small 
numbers of photons detected in the short time intervals at $\gamma$-ray energies. The 
shortest time-scale variations detected for blazars with EGRET are for PKS 1622-297 
(Mattox et al. 1997b) and 3C 279 (Wehrle et al. 1997). For both these objects the flux 
was found to increase by a factor of two or more in less than 8 hours. Other objects 
that have shown flux variations over the period of a few days are 3C 279 (Kniffen et al. 
1993), 3C 454.3 (Hartman et al. 1993), 4C 38.41 
(Mattox et al. 1993), PKS 1406-076 (Wagner et al. 1995), and PKS 0528+134 
(Hunter et al. 1993; Mukherjee et al. 1996). The short time-scale of $\gamma$-ray 
flux variability (e.g. in 3C 279 or PKS 1622-297) when combined with the large inferred 
$\gamma$-ray luminosities, implies that the blazar emission region is very compact. 
Gamma-ray tests for beaming from variability and flux measurements using the 
Elliott-Shapiro relation and $\gamma$-ray transparency arguments 
are summarized in a recent review on $\gamma$-ray blazars by 
Hartman et al. (1997). A factor-of-two flux variation on an observed 
time-scale $\delta t_{\rm obs}$ limits the size $r$ of a stationary 
isotropically emitting region to be roughly $r\leq c\delta t_{\rm obs}/(1+z)$ 
by simple light-travel time arguments. 
Under the assumptions of isotropic radiation and Eddington-limited accretion, the 
implied minimum black hole masses of blazars are $\geq 8\times 10^{11}$ $M_\odot$ 
for PKS 1622-297 (Mattox et al. 1997b) from EGRET observations and 
$\geq 7.5\times 10^{8}$ $M_\odot$ for PKS 0528+134 from COMPTEL observations 
(Collmar et al. 1997). 

\section{Luminosity}

The $\gamma$-ray luminosity can be estimated by considering the relationship 
between the 
observed differential energy flux $S_0(E_0),$ where the subscript ``$0$'' 
denotes the observed or present value, and $Q_e$ the power emitted in $dE$, 
where $E=E_0(1+z)$ in the Friedman universe. 
$$Q_e[E_0]=4\pi S_0(E_0)(1+z)^{b-1}\Theta {D_L}^2(z,q_0)\eqno (1)$$
where 
$$D_L={c\over{H_0{q_0}^2}}\bigl[ 1-q_0+q_0z+(q_0-1)(2q_0z+1)^{1/2}\bigr]\equiv{{cz}\over{H_0}}g(z,q_0).\eqno (2)$$
\noindent
$H_0$ is the Hubble parameter, $q_0$ is the deceleration parameter, $b$ is the 
spectral index, $z$ is the redshift, and 
$\Theta$ is the beaming factor. $H_0$ is chosen to be 70 and $q_0$ to be 0.5, 
although the results obtained here are not highly sensitive to these choices. 
The beaming factor is taken to be 1. The spectral index is obtained using the 
analysis described in \S 4. The luminosity as a function of the redshift 
is determined using equation (1) and is plotted in Figure 4 for the 
blazars detected by EGRET. The typical detection threshold for EGRET as a 
function of $z$, for relatively good conditions, is also shown in the 
figure. The actual threshold varies somewhat with exposure and region of the 
sky, and the average threshold is a little higher than the curve shown, but 
the shape is the same. The BL Lac objects are indicated in the figure 
with dark diamonds, and one sees clearly that they are predominantly closer 
and lower in luminosity. 

\begin{figure}[t!] 
\centerline{\epsfig{file=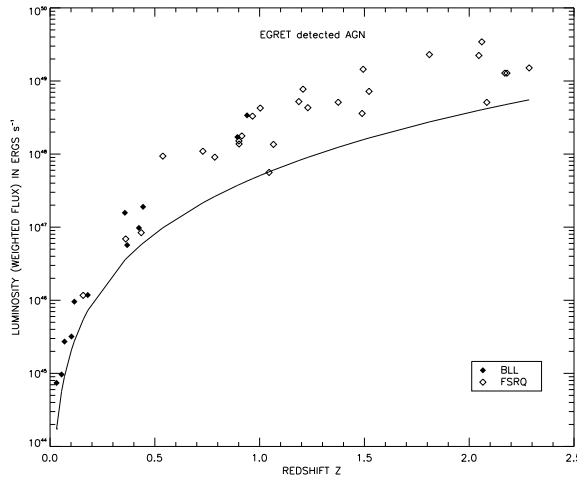,height=3.0in,bbllx=70pt,bblly=720pt,bburx=650pt,bbury=1300pt,clip=.}}
\vspace{10pt}
\caption{Luminosity vs redshift for blazars detected by EGRET. The BL Lac 
objects are indicated with filled symbols. The typical detection threshold for 
EGRET is shown as a solid curve.}
\label{fig4}
\end{figure}
 
Recently, Chiang \& Mukherjee (1998) have calculated the evolution and 
luminosity function of the EGRET blazars, and have estimated the contribution 
of this source class to the diffuse extragalactic gamma-ray background. They 
find that the evolution is consistent with pure luminosity evolution. 
According to their estimates, only 25\% of the diffuse extragalactic emission 
measured by SAS-2 and EGRET can be attributed to unresolved $\gamma$-ray 
blazars, contrary to some of the other estimates (eg. Stecker \& Salamon 1996).
Below 10 MeV, the average blazar spectrum suggests that only about 50\% of the 
measured $\gamma$-ray emission could arise from blazars (Sreekumar, Stecker 
\& Kappadath 1997). This leads to the exciting possibility that other sources 
of diffuse extragalactic $\gamma$-ray emission must exist. 

\section{Spectra}

\subsection{Spectra in the EGRET energy range}

EGRET spectra of blazars typically covers at least two decades in energy (from 30 MeV 
to 10 GeV) and are well described by a simple power-law model of the form 
$F(E)=k(E/E_0)^{-\alpha}$ photons cm$^{-2}$ s$^{-1}$ MeV$^{-1}$, 
where the photon spectral index, $\alpha$, and the coefficient, $k$, 
are the free parameters. The energy normalization factor, $E_0$, is 
chosen so that the statistical errors in the power law index and the overall 
normalization are uncorrelated. 

The average blazar spectrum has a spectral index of about $-2.15$. Figure 5 shows 
the photon spectral index of the blazars plotted as a function of the redshift. There 
are marginal indications that suggest that the BL Lac objects have slightly harder 
spectrum 
in the EGRET energy range than the FSRQs. Mukherjee et al. (1997) find that the average 
spectral index of the BL Lac objects is about $-2.03$, compared to about $-2.20$ for 
the FSRQs. For some individual blazars there has been noted a trend for the spectrum 
to harden during a flare state (eg. in blazars 1222+216, 1633+382, and 0528+134; 
Sreekumar et al. 1996; Mukherjee et al. 1996). 
A spectral study of blazars as a class has been performed by 
M\"ucke et al. (1996) and Pohl et al. (1997). 

\begin{figure}[t!] 
\centerline{\epsfig{file=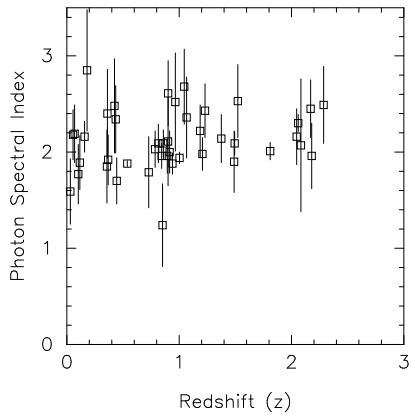,height=2.5in,bbllx=30pt,bblly=140pt,bburx=500pt,bbury=650pt,clip=.}}
\vspace{10pt}
\caption{Photon spectral index as a function of the redshift for blazars 
detected by EGRET.}
\label{fig5}
\end{figure}
 
\subsection{Spectral Energy Distributions and Gamma-ray Models}
 
The processes by which $\gamma$-rays are produced in blazars can be best understood by 
the study of the correlated multiwaveband observations of blazars extending from radio 
to $\gamma$-ray wavebands. One of the most significant findings of EGRET is that, in the radio to $\gamma$-ray 
multiwavelength spectra of blazars, the power in the $\gamma$-ray range equals or 
exceeds the power in the infrared-optical band. Any model of high-energy $\gamma$-ray 
emission in blazars needs to explain this basic observational fact. The high $\gamma$-ray 
luminosity of the blazars suggests that the emission is likely to be beamed and, 
therefore, Doppler-boosted into the line of sight. This is in agreement with the strong 
association of EGRET blazars with radio-loud flat-spectrum radio sources, with 
many of them showing superluminal motion in their jets. This information has 
helped to favor jet models of emission over models in which the $\gamma$-ray production 
is directly associated with accretion onto a massive black hole 
(e.g. Becker \& Kafatos 1993). 

The jet models explain the radio to UV continuum from blazars as synchrotron 
radiation from high energy electrons in a relativistically outflowing jet which 
has been ejected from an accreting supermassive black hole (Blandford \& K\"onigl 1979). 
The emission in the MeV-GeV range is believed to be due to the 
inverse Compton scattering of low-energy photons by the same 
relativistic electrons in the jet. However, two main issues remain questionable: 
the source of the soft 
photons that are inverse Compton scattered, and the structure of the inner jet, 
which cannot be imaged directly. The soft photons can originate as 
synchrotron emission either from within the jet (the synchrotron-self-Compton 
or SSC process: Maraschi, Ghisellini, \& Celotti 1992; Bloom \& Marscher 1996), 
or from a nearby accretion disk, or they can be disk radiation reprocessed 
in broad-emission-line clouds (the external radiation Compton process or 
the ERC process: Dermer \& Schlickeiser 1994; Sikora, Begelman, \& Rees 1994; 
Blandford \& Levison 1995; Ghisellini \& Madau 1996). In contrast to these 
leptonic jet models, the proton-initiated cascade (PIC) model 
(Mannheim \& Biermann 1989, 1992) predicts that the high-energy emission 
comes from knots in jets as a consequence of diffusive shock acceleration of 
protons to energies so high that the threshold of secondary particle production 
is exceeded. 

Figure 6 shows the simultaneous spectral energy distribution 
of 3C 279 during January-February 1996, when the source was detected at its 
highest state ever (Wehrle et al. 1997). The figure shows the relative amounts of 
energy detected in equal logarithmic frequency bands. The power output in $\gamma$-rays 
dominates the bolometric luminosity of the sources, as mentioned in \S 1. Wehrle et al. 
(1997) note that the $\gamma$-rays vary by more than the square of the observed 
IR-optical flux change, a fact that could be hard to explain by some specific blazar 
emission models. Although the data do not rule out SSC models, Wehrle et al. point out 
that the data are most likely explained by the ``mirror'' model of 
(Ghisellini \& Madau 1996). 
In this model the flaring region in the jet photo-ionizes nearby 
broad-emission-line clouds, which in turn provide low energy external seed photons that 
are inverse Compton-scattered to high energy $\gamma$-rays. 

\begin{figure}[t!] 
\centerline{\epsfig{file=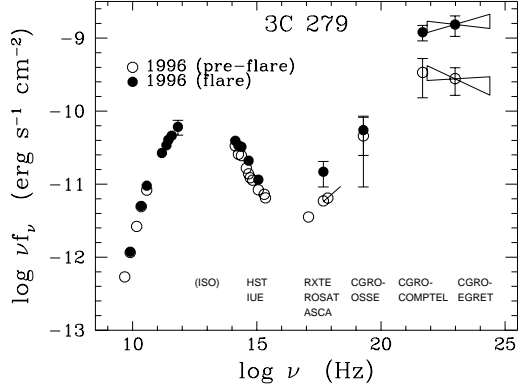,height=2.5in,bbllx=650pt,bblly=40pt,bburx=1300pt,bbury=650pt,clip=.}}
\vspace{10pt}
\caption{Radio to $\gamma$-ray energy distribution of 3C 279 in low (open circles) 
and flaring state (filled circles) in 1996 January-February (Wehrle et al. 
1997).}
\label{fig6}
\end{figure}
 
Recently, a model combining the ERC and SSC scenarios has been used to fit the 
simultaneous COMPTEL and EGRET spectra of PKS 0528+134 by B\"ottcher \& 
Collmar (1998). 
Figure 7 shows their fit to the gamma-ray spectrum during the high 
state of the source during March 1993. In their model B\"ottcher \& Collmar 
assume a spherical blob filled with ultrarelativistic pair plasma which is 
moving out along an existing jet structure perpendicular to an accretion 
disk around a black hole of mass $5\times 10^{10}\ M_\odot$. They argue that 
the observed spectral break between COMPTEL and EGRET energy ranges can 
plausibly be explained by a variation of the Doppler beaming factor in the 
framework of a relativistic jet model for AGNs.  

The EGRET results have demonstrated that in order 
to model the spectra of blazars it is  very important to get a truly 
simultaneous coverage across  the entire electromagnetic spectrum before, during, and 
after a flare in the high-energy $\gamma$-ray emission. 
The limited data that we have on most of the blazars prevents us from being able 
to distinguish between the different theoretical models, on the basis of the 
spectra alone.  For example, both the SSC and ERC models have been shown to reproduce the 
multiwavelength spectrum of 3C 279 rather well (Hartman et al.  1996b; Maraschi, 
Ghisellini \& Celotti 1992; Ghisellini \& Maraschi 1996). The SSC model was 
similarly found 
to fit the multiwavelength spectrum of PKS 0528+134 during the March 1993 
flare reasonably well (Mukherjee et al.  1996). The low-state data of PKS 0528+134 (Aug 1994) 
was fit well with the ERC model, as demonstrated by Sambruna et al.  (1997). 
The SSC, ERC, and PIC models have all been 
shown to fit the multiwavelength spectrum of 3C 273 well (von Montigny 1997). 

\begin{figure}[t!] 
\centerline{\epsfig{file=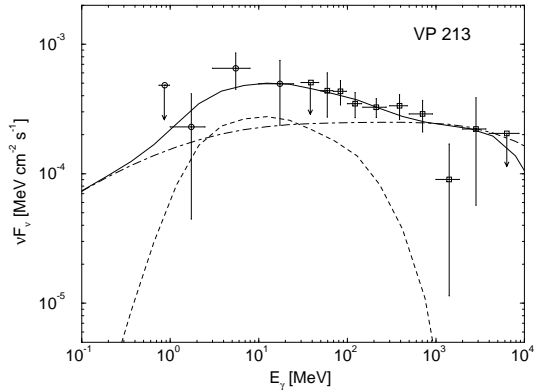,height=2.5in,bbllx=-50pt,bblly=-570pt,bburx=800pt,bbury=0pt,clip=.}}
\vspace{10pt}
\caption{Fit to the $\gamma$-ray (COMPTEL and EGRET) spectrum of PKS 0528+134 
in its high state in VP 213. See B\"ottcher \& Collmar (1998) for fit 
parameters.}
\label{fig7}
\end{figure}
 
The differences between the $\gamma$-ray variability properties of BL Lac objects 
and FSRQs can be explained in light of the model of Ghisselini and Madau (GM) (1996). 
In their model, soft photons from the jet are reprocessed by broad line region (BLR) 
clouds. Subsequently, these soft photons are emitted back into the jet where 
they scatter off of electrons in relativistically moving  ``blobs'' to create high-energy 
$\gamma$-rays. Since BL Lac objects generally have very weak emission lines, it may be
that they have much less BLR gas available for reprocessing soft
photons. If there is an initial outburst of soft photons created via the
synchrotron process in a jet ``blob,'' then BL Lac objects can still
create $\gamma$-rays via the SSC process, though perhaps with lower amplitude
than $\gamma$-rays created via the GM model. (This effect is somewhat dependent on 
adjustable model parameters.) In this scenario,
BL Lac objects may undergo several SSC outbursts which fail to reach the
EGRET detection threshold, thus giving the appearance that BL Lac objects 
as a class experience less dramatic and less frequent $\gamma$-ray
flares. The general properties of the low 
frequency outbursts of BL Lac objects, however, would be very similar to that of the 
FSRQs.

In order to achieve a better understanding of the emission mechanisms of 
$\gamma$-rays from blazars a study of the correlated short time scale 
($\sim 1-3$ days)  $\gamma$-ray variations with those at other frequency bands 
is needed. Since the predictions of time delays between the flux changes at 
various frequencies are different for the individual models for both the seed 
photons and the nature of the inner jet, this method could provide a means to 
discriminate between the different models. The differences in the 
model predictions are discussed in more detail by Marscher et al. (1995). 
Gamma-ray variability in the different models may have different impacts on 
the spectral behavior during the build-up and decline of an outburst. 
Studying the short-time-scale behavior and looking for spectral changes while 
following a complete outburst may be the key to pin down the basic emission 
mechanisms. 

\section{Summary}

In conclusion, the EGRET results have shown the importance of the $\gamma$-ray 
window on blazars. The high luminosities and strong time 
variability observed have pushed theoretical models to emphasize 
relativistic jets of particles seen at small angles to the line of 
sight.  The EGRET observations have established that the $\gamma$-ray window 
is critical for understanding the properties of blazars. Future observations 
with CGRO and successor $\gamma$-ray observatories like INTEGRAL and GLAST 
should play a key role in resolving the physics of these powerful sources. 

The author presents this work on behalf of the EGRET Team and acknowledges 
contributions from D. L. Bertsch, S. D. Bloom, 
B. L. Dingus, J. A. Esposito, C. E. Fichtel, 
R. C. Hartman, S. D. Hunter, G. Kanbach, 
D. A. Kniffen, Y. C. Lin, H. A. Mayer-Hasselwander, 
L. M. McDonald, P. F. Michelson, C. von Montigny, 
A. M\"ucke, P. L. Nolan, M. Pohl, O. Reimer, 
E. Schneid, P. Sreekumar, and D. J. Thompson. 
The author would particularly like to thank P. Sreekumar for critical comments 
on the draft. The author also acknowledges support from NASA Grant 
NAG5-3696. 

\section{References}

\noindent
Becker, P. A. \& Kafatos, M. 1993, in: Proceedings of the 2nd COMPTON 
Symposium, College Park, MD 1993, AIP Conference Proc. No. 304, eds: 
C. E. Fichtel, N. Gehrels, \& J. P. Norris, pg. 620

\noindent
Bertsch, D.  L., et al. 1989, {\sl Proc. of the Gamma Ray Observatory 
Science Workshop}, ed. W. N. Johnson, 2, 52

\noindent
Blandford, R. D. \& K\"onigl, A. 1979, ApJ, 232, 34

\noindent
Blandford, R. D. \& Levison, A. 1995, ApJ, 441, 79

\noindent
Bloom, S. D. \& Marscher, A. P. 1996, ApJ, 461, 657

\noindent
Bloom, S. D., et al. 1997, ApJ, 490, L145

\noindent
B\"ottcher, M. \& Collmar, W. 1998, A\&A, 329, L57

\noindent
Chiang, J. \& Mukherjee, R. 1998, ApJ, 496, 752

\noindent
Collmar, W., et al. 1997, A\&A, 328, 33

\noindent
Dermer, C. D. \& Schlickeiser, R. 1994, ApJS, 90, 945

\noindent
Esposito, J. A., et al. 1998, in preparation

\noindent
Ghisellini, G. \& Madau, P. 1996, MNRAS, 280, 67

\noindent
Ghisellini, G. \& Maraschi, L. et al. 1996, ``Blazar Continuum Variability,'' 
A. S. P. Conf. Series Vol. 110, pg 436

\noindent
Hartman, R. C., et al. 1993, ApJ, 407, L41

\noindent
Hartman, R. C., et al. 1996a, ``Blazar Continuum Variability,'' A. S. P. 
Conf. Series Vol. 110, pg 333

\noindent
Hartman, R. C., et al. 1996b, ApJ, 461, 698

\noindent
Hartman, R. C., et al. 1997, {\sl Proc. of the Fourth Compton Symposium}, 
eds. C. D. Dermer, M. S. Strickman, \& J. D. Kurfess, CP410, 307

\noindent
Hartman, R. C., et al. 1998, ApJ, submitted

\noindent
Hughes, E. B., et al. 1980, IEEE Trans. Nucl. Sci., NS-27, 364

\noindent
Hunter, S. D., et al. 1993, ApJ, 409, 134

\noindent
Hunter, S. D., et al. 1997, ApJ, 481, 205

\noindent
Kanbach, G., et al. 1988, Space Sci. Rev., 49, 69

\noindent
Kanbach, G., et al. 1989, {\sl Proc. of the Gamma Ray Observatory Science 
Workshop}, ed. W. N. Johnson, 2, 1

\noindent
Kniffen, D. A., et al. 1993, ApJ, 411, 133

\noindent
Mannheim, K. \& Biermann, P. L. 1989, A\&A, 221, 211

\noindent
Mannheim, K. \& Biermann, P. L. 1992, A\&A, 53, L21

\noindent
Maraschi, L., Ghisellini, G., \& Celotti, A. 1992, ApJ, 397, L5

\noindent
Marscher, A. P., et al. 1995, PNAS, 92, 11439

\noindent
Mattox, J. R., et al. 1993, ApJ, 410, 609

\noindent
Mattox, J. R., et al. 1996, ApJ, 461, 396

\noindent
Mattox, J. R., et al. 1997a, ApJ, 481, 95

\noindent
Mattox, J. R., et al. 1997b, ApJ, 476, 692

\noindent
McLaughlin, M. A., et al. 1996, ApJ, 473, 763

\noindent
von Montigny, C., et al. 1995, ApJ, 440, 525

\noindent
von Montigny, C., et al. 1997, ApJ, 483, 161

\noindent
M\"ucke A., et al. 1996, IAU Symposium 175, (Dordrecht: Kluwer)

\noindent
Mukherjee, R., et al. 1996, ApJ, 470, 831

\noindent
Mukherjee, R., et al. 1997, ApJ, 490, 116

\noindent
Pohl, M., et al. 1997, A\&A, submitted

\noindent
Sambruna, R. M., et al. 1997, ApJ, 474, 639

\noindent
Sikora, M., Begelman, M. C., \& Rees, M. J. 1994, ApJ, 421, 153

\noindent
Sreekumar, P., et al. 1996, ApJ, 464, 628

\noindent
Sreekumar, P., Stecker, F. W., \& Kappadath, S. C. 1997, {\sl Proc. of the 
Fourth Compton Symposium}, eds. C. D. Dermer, M. S. Strickman, \& 
J. D. Kurfess, CP410, 307

\noindent
Sreekumar, P., et al. 1998, ApJ, 494, 523

\noindent
Stecker, F. W., \& Salamon, M. H. 1996, ApJ, 464, 600

\noindent
Swanenburg, B. N., et al. 1978, Nature, 275, 298

\noindent
Thompson, D. J., et al. 1993a, ApJS, 86, 629

\noindent
Thompson, D. J., et al. 1995, ApJ, 101, 259

\noindent
Thompson, D. J., et al. 1996, ApJS, Vol. 107, 227

\noindent
Wagner, S. et al. 1995, ApJ, 454, L97

\noindent
Wehrle, A. et al. 1997, ApJ, submitted

\end{document}